\documentclass[twocolumn,showpacs,amsmath,amssymb,prl,superscriptaddress,floatfix]{revtex4}

\usepackage{graphicx}
\usepackage{dcolumn}
\usepackage{bm}
\usepackage{xcolor}
\definecolor{red}{rgb}{0.7,0,0}
\definecolor{green}{rgb}{0.,0.35,0.}
\definecolor{blue}{rgb}{0.2,0.2,0.7} 
\definecolor{black}{rgb}{0.15,0.15,.15}
\def\gapx{\lower 2pt \hbox{$\buildrel>\over{\scriptstyle{\sim}}$\ }}
\def\lapx{\lower 2pt \hbox{$\buildrel<\over{\scriptstyle{\sim}}$\ }}

\usepackage{graphicx}
\usepackage{dcolumn}
\usepackage{bm}

\include{revmacro}

\begin{document}

\title{Supersolid droplet crystal in a dipole-blockaded gas}

\author{F. Cinti}
\affiliation{Department of Physics, University of Alberta, Edmonton, Alberta, Canada T6G 2J1}
\author{P. Jain}
\affiliation{Department of Physics, University of Alberta, Edmonton, Alberta, Canada T6G 2J1}
\author{M. Boninsegni}
\affiliation{Department of Physics, University of Alberta, Edmonton, Alberta, Canada T6G 2J1}
\author{A. Micheli}
\affiliation{IQOQI and Institute for Theoretical Physics, University of Innsbruck, 6020 Innsbruck, Austria.}
\author{P. Zoller}
\affiliation{IQOQI and Institute for Theoretical Physics, University of Innsbruck, 6020 Innsbruck, Austria.}
\author{G. Pupillo}
\affiliation{IQOQI and Institute for Theoretical Physics, University of Innsbruck, 6020 Innsbruck, Austria.}
\date{\today}
\begin{abstract}
A novel supersolid phase is predicted for  an ensemble of Rydberg atoms in the dipole-blockade regime, interacting via a repulsive dipolar potential  ``softened" at short distances. Using exact numerical techniques, we study the low temperature phase diagram of this system, and observe an intriguing phase consisting of a crystal of mesoscopic superfluid droplets. At low temperature, phase coherence throughout the whole system, and the ensuing bulk superfluidity, are established through tunnelling of identical particles between neighbouring droplets.
\end{abstract}



\maketitle

The search for novel phases of matter drives much of the current research in condensed matter physics. Of particular interest are phases simultaneously displaying different types of order. A chief example, of great current interest, is the so-called {\em supersolid}, namely a phase featuring crystalline order, and also capable of sustaining dissipation-less flow. Attempts to observe experimentally a supersolid phase of matter, primarily in a crystal of solid helium, have spanned four decades since early theoretical predictions \cite{Andreev69}. The most credible claim of such an observation to date \cite{kim04a,kim04b}, has been subjected to in-depth scrutiny over the past few years, and it seems fair to state that agreement is lacking at the present time, as to whether experimental findings indeed signal a supersolid phenomenon \cite{nikolay07}.
\\ \indent
A new, fascinating avenue to the observation of supersolid and  other phases of matter not yet observed (or even thought of), is now opened by advances in cold atom physics,  providing not only remarkably clean and controlled experimental systems, but also allowing one to ``fashion" artificial inter-particle potentials, not arising in any known condensed matter system. This allows one to address a key theoretical question, namely which two-body interaction potential(s), if any, can lead to the occurrence of a supersolid phase in free space (i.e., not on a lattice).
\\ \indent
\\ \indent
In a recent article  \cite{Henkel10},  Henkel \textit{et al.} have proposed, based on a mean-field treatment, that a Bose 
condensate of particles interacting through an effective potential which flattens off at short distance, might support a density modulation.
In this Letter,  we show by first principle numerical simulations that interaction potentials which combine a long-distance repulsion with a short-distance cutoff, lead in fact to the appearance of a {\it novel} self-assembled crystalline phase of mesoscopic superfluid droplets in a system of bosons. Furthermore, such a crystal can turn supersolid in the $T\to 0$ limit, as tunneling of particles across neighbouring droplets takes place, and superfluid phase coherence is established across the whole system, as individual separate Bose condensates (droplets) organize into a single, global condensate. 
Specifically, we consider the following two-body potential:
\begin{equation}
\label{pot1}
v(r) = \left\{
\begin{array}{rl} D/a^3 & \text{if } r \le a\, \\
\\
D/r^3 & \text{if } r> a\,
\end{array}
\right., 
\end{equation}
$D$ being the characteristic strength of the interaction.  This kind of interaction potential can be realized with cold dipole-blockaded Rydberg atoms \cite{Lukin2001,Tong2004,Vogt2006,Heidemann2008,pritchard2010}. The parameters $D$ and $a$ above can be controlled with external fields \cite{Pupillo10,Henkel10} (we come back to this point below). 

Our system of interest comprises $N$ identical bosons of mass $m$, confined to two dimensions \footnote{Confinement to two dimensions can be achieved using a tight (magnetic or optical) trap along $z$.}.  The many-body Hamiltonian is the following (in dimensionless form):
\begin{equation}
\label{ham2}
{{\cal H}} =
-\frac{1}{2}\sum_{i=1}^{N}\nabla^2_i
+\sum_{i>j}{v}(r_{ij})
\end{equation}
where $r_{ij}=|{\bf r}_i-{\bf r}_j|$ is the distance between particles $i$ and $j$, and $v$ is given by Eq. (\ref{pot1}). All lengths are expressed in terms of the characteristic length $r_\circ=mD/\hbar^2$, and we introduce a dimensionless cutoff $R_c = a/r_\circ$ for the  potential (\ref{pot1}).  The system is enclosed in a square cell of area $A$, with periodic boundary conditions. The particle density is $n=N/A$, but we shall express our results in terms of the (dimensionless) inter-particle distance $r_s$=$1/\sqrt{nr_\circ^2}$. The energy scale is $\epsilon_\circ=D/r_\circ^3=\hbar^2/mr_\circ^2$.
\\ \indent
The low-temperature phase diagram of such a system has been explored by means of first principles numerical simulations, based on the Continuous-space Worm Algorithm  \cite{Boninsegni2006L,Boninsegni2006}.
Numerical results shown here pertain to simulations with a number of particles $N$ varying between 50 and 400, in order to carry out extrapolation of the results to the thermodynamic limit. Our ground state estimates are obtained as extrapolations of  results at finite temperature. Details of the simulations are standard, as the use of the potential (\ref{pot1}) entails no particular technical difficulty.
\\ \indent
In the limit $R_c \ll r_s$, the truncation of the dipolar potential at short distances does not play an important role, and the low temperature phase diagram of (\ref{ham2}) is that of purely dipolar bosons in two dimensions, investigated previously by several authors \cite{Buchler2007,Astrakharchik2007}. It is known that for $r_s$ $\lapx$ $r_s^C$ = 0.06 the ground state of the system is a triangular crystal, whereas for $r_s$ $\gapx$ $r_s^L$ =  0.08 it is a uniform superfluid (in the intermediate density range a more complex scenario is predicted \cite{spivak04}). As we show below, a very different physics  sets in when $R_c\ \gapx r_s$, in the density ranges which correspond to either the crystalline or superfluid phase in the purely dipolar system.
\begin{figure}[!t]
\begin{center}
\includegraphics[width=0.45\textwidth]{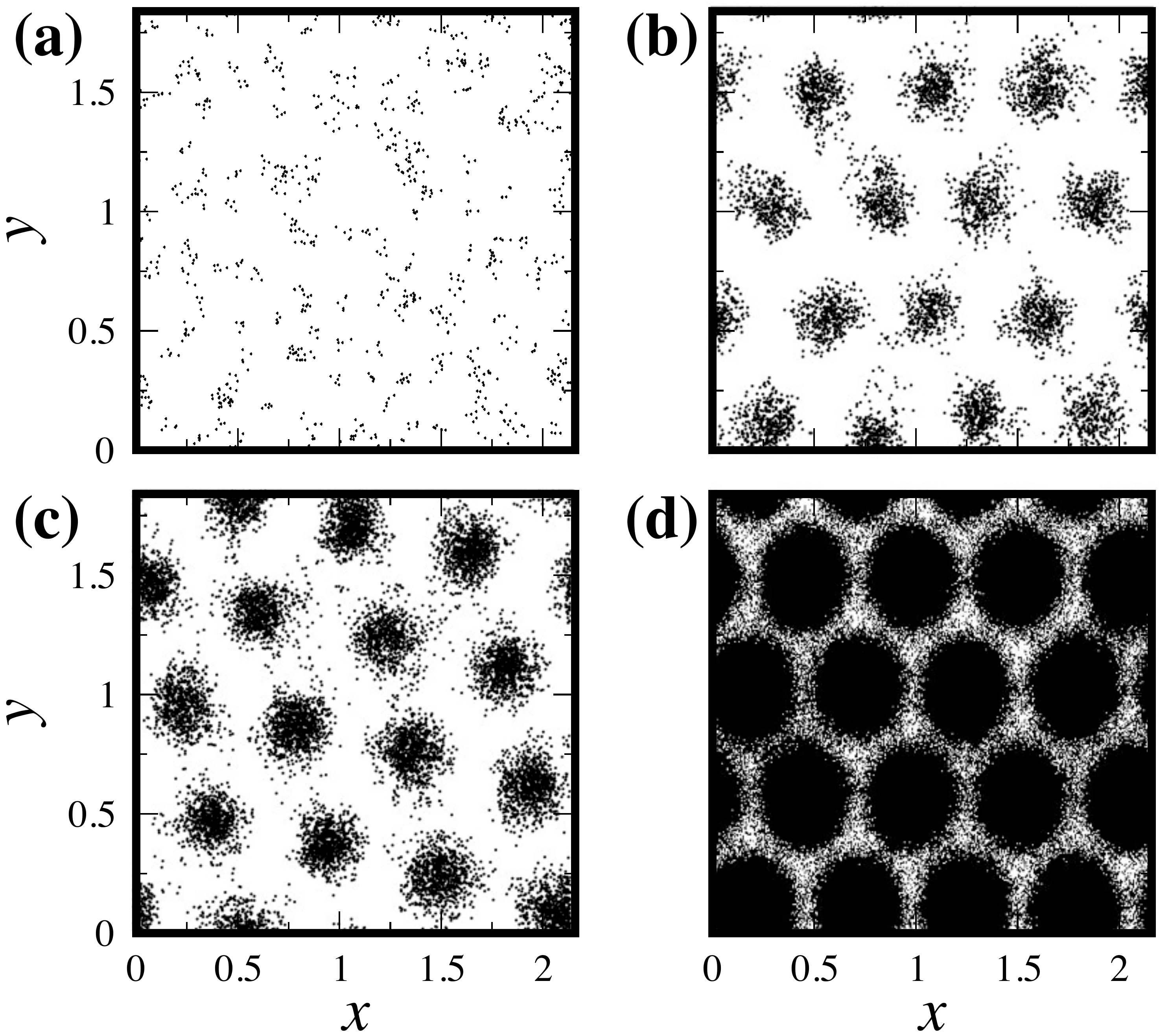}
\caption{Snapshots of a system of bosons  interacting via potential (\ref{pot1}), at the four different temperatures  200 (a), 20 (b), 1.0 (c) and 0.1 (d), expressed in units of $\epsilon_\circ$. Points shown are taken along individual particle world lines. The nominal value of $r_s$ in this case is 0.14, whereas the cutoff of the potential (\ref{pot1}) is $R_c$=0.3.
}
\label{default}
\end{center}
\end{figure}
\begin{figure}[t!]
\begin{center}
\includegraphics[width=0.38\textwidth]{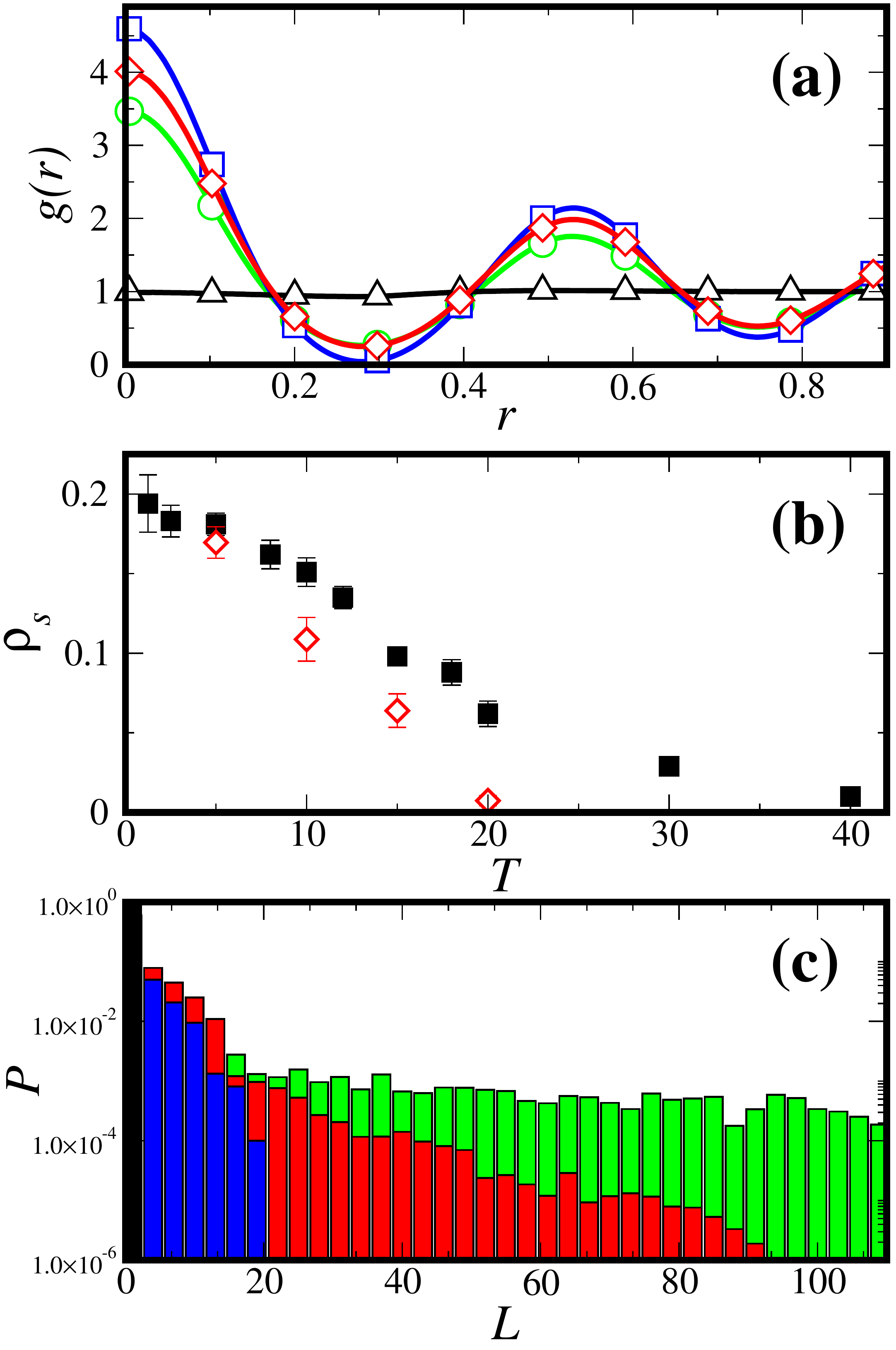}
\caption{(Color online) 
Results shown are for $r_s$=0.14 and $R_c$=0.3.
Temperature is in units of $\epsilon_\circ$.
(a): Pair correlation function $g(r)$ at a temperature $T$= 200 (triangles), 20 (squares), 1.0 (diamonds) and 0.1 (circles). The simulated system comprises $N$=200 particles.
(b): Superfluid density \text{vs.} $T$ for systems with $N$=100 (square), and 200 (diamond) particles.
(c): Frequency of occurrence of  permutation cycles of length $L$ at the same four temperatures reported in panel (a). Longer permutation cycles occur at lower temperature.}
\label{cycles}
\end{center}
\end{figure}
Fig. \ref{default} shows typical configurations (i.e., particle world lines) produced by Monte Carlo simulations of a system of bosons interacting via the potential \eqref {pot1}, at a nominal density corresponding to $r_s=0.14$, at different temperatures  spanning three orders of magnitude. The value of the cutoff $R_c$ in this case is 0.3.
At the highest temperature,  a simple classical gas phase is observed, as shown by the pair correlation function $g(r)$, shown in Fig. \ref{cycles} (a), which is just a constant (note that $g(r)$ does not vanish at the origin, owing to the flattening off of the potential at short distance). As $T$ is decreased,
an intriguing effect takes place, namely particles bunch into mesoscopic droplets, in turn forming a regular (triangular) crystal.  This is shown qualitatively in the snapshots in Fig. \ref{default}, but also confirmed quantitatively by the structure of the $g(r)$ as well (Fig. \ref{cycles}(a)), which displays pronounced, broad maxima, as well as well-defined minima, where the function approaches zero. We henceforth refer to this phase as the \textit{droplet-crystal} phase. \\ \indent
The formation of such droplets is a purely classical effect, that depends on the flattening off of the repulsive inter-particle potential below the cutoff distance. In fact, a simple estimate of the number  $N_d$ of particles per droplet, can be obtained by considering a triangular lattice of point-like dipoles, each one of strength $\propto N_d$ (as it comprises $N_d$ particles), and by minimizing with respect to $N_d$  the potential energy per particle, for a fixed density. The result is
\begin{equation}
\label{mestimate}
N_d=\gamma \biggl (\frac{R_c}{r_s} \biggr ) ^2
\end{equation}
 where $\gamma \approx 2.79$. Eq. \eqref{mestimate} furnishes a fairly accurate estimate of $N_d$  for the (wide) range of values of the parameters $r_s$ and $R_c$ explored here. For instance, using the parameters of Fig.~\ref{default},
we find from \eqref{mestimate} $N_d\approx13$, which agrees quite well with our simulation result. It is worth noting that a similar sort of pattern formation, due to competing interactions, has been previously established for classical colloidal systems \cite{Liu2008,Archer2008}.
\\ \indent
In the $T\to0$ limit, long exchanges of identical particles can take place, as a result of particles tunneling from one droplet to an adjacent one. Long exchanges of particles can result in a finite superfluid response throughout the whole system, and indeed for $R_c$ $\gapx r_s^L$  we observe such a bulk superfluid signal, in a range of values of  $r_s$ in the vicinity of $R_c/2$. A typical result is shown in Fig. \ref{cycles}(b) \footnote{We compute directly the {\em bulk} superfluid fraction through the usual {\it winding number} estimator for the bulk system.}.
Because superfluidity arises in concomitance with the droplet-crystal structure, 
the denomination {\em supersolid} seems appropriate. 

In order to establish that droplets are individually superfluid, one may consider the statistics of permutation cycles. Fig. \ref{cycles} (c) shows the frequency of occurrence of exchange cycles involving a varying number $L$ of particles ($1 \le L \le N$), at three different temperatures, at the physical conditions of Fig. \ref{default}. As one can see, as the temperature is lowered exchange cycles involving growing numbers of particles occur. At low temperature they involve almost all the particles in the system; however, even at a higher temperature (e.g., $T$=20 in Fig. \ref {cycles}(c)) one observes exchanges comprising a number of particles up to $\sim N_d$, i.e., particles inside an individual droplet. This is evidence that droplets are individually Bose condensed and superfluid, even though the system as a whole does not display superfluidity. 
That droplets should be superfluid at low $T$ is not surprising, given that particles in a droplet are essentially non-interacting, due to the flatness of the potential at short distance. However, that droplets are themselves superfluid does not imply that a bulk supersolid phase will {\it always} occur in the $T$ $\to$ 0 limit, as discussed above.
\\ \indent
\begin{figure}[!t]
\begin{center}
\includegraphics[width=0.4\textwidth]{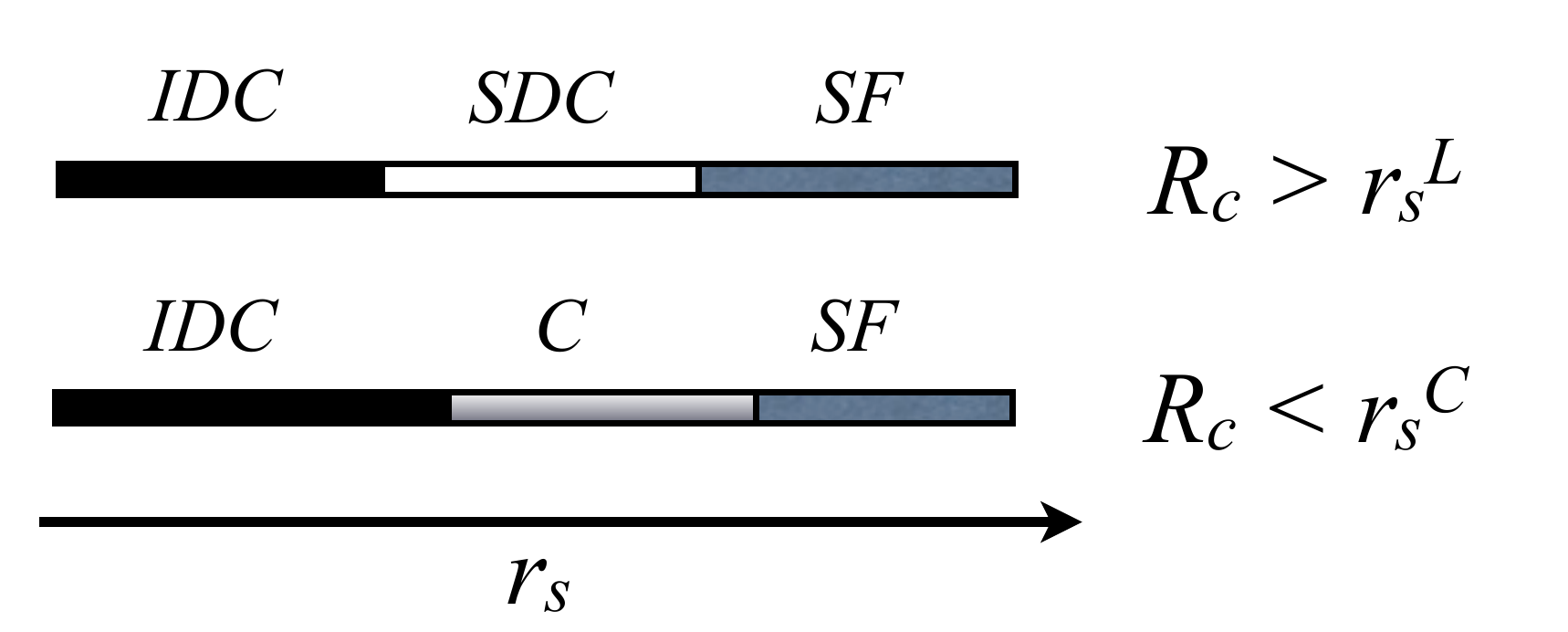}
\caption{ Schematic ground state phase diagram of \eqref {ham2} as a function of $r_s$. The superfluid droplet crystal (SDC) is sandwiched between an insulating droplet crystal (IDC) and a superfluid (SF). For $R_c$ $\lapx$ $r_s^C$, an IDC, a single-particle crystal (C) and a superfluid phases are observed. The widths of the SDC and C regions  depend on the value of $R_c$. }
\label{phase}
\end{center}
\end{figure}
\indent
At $T$ = 0, the supersolid phase is sandwiched between an insulating droplet crystal at high density (i.e., lower $r_s$) and a homogeneous superfluid phase at lower density. For $R_c$ $\lapx$ $r_s^C$, only two insulating phases are observed,  namely the insulating droplet crystal at high density and the crystal of single particles, already detected in Refs. \cite{Buchler2007,Astrakharchik2007}, as well as a superfluid phase at lower density.
All of this is summarized in the schematic phase diagram shown in Fig. \ref{phase}.
It is important to stress that supersolid behaviour in this system 
originates from tunnelling of particles between droplets which are themselves individually superfluid, so that the individual superfluid droplets connect to form a bulk superfluid. This is reminiscent of the phase-locking mechanism in a (self-assembled) array of Josephson junctions.

\begin{figure}[!t]
\begin{center}
\includegraphics[width=0.45\textwidth]{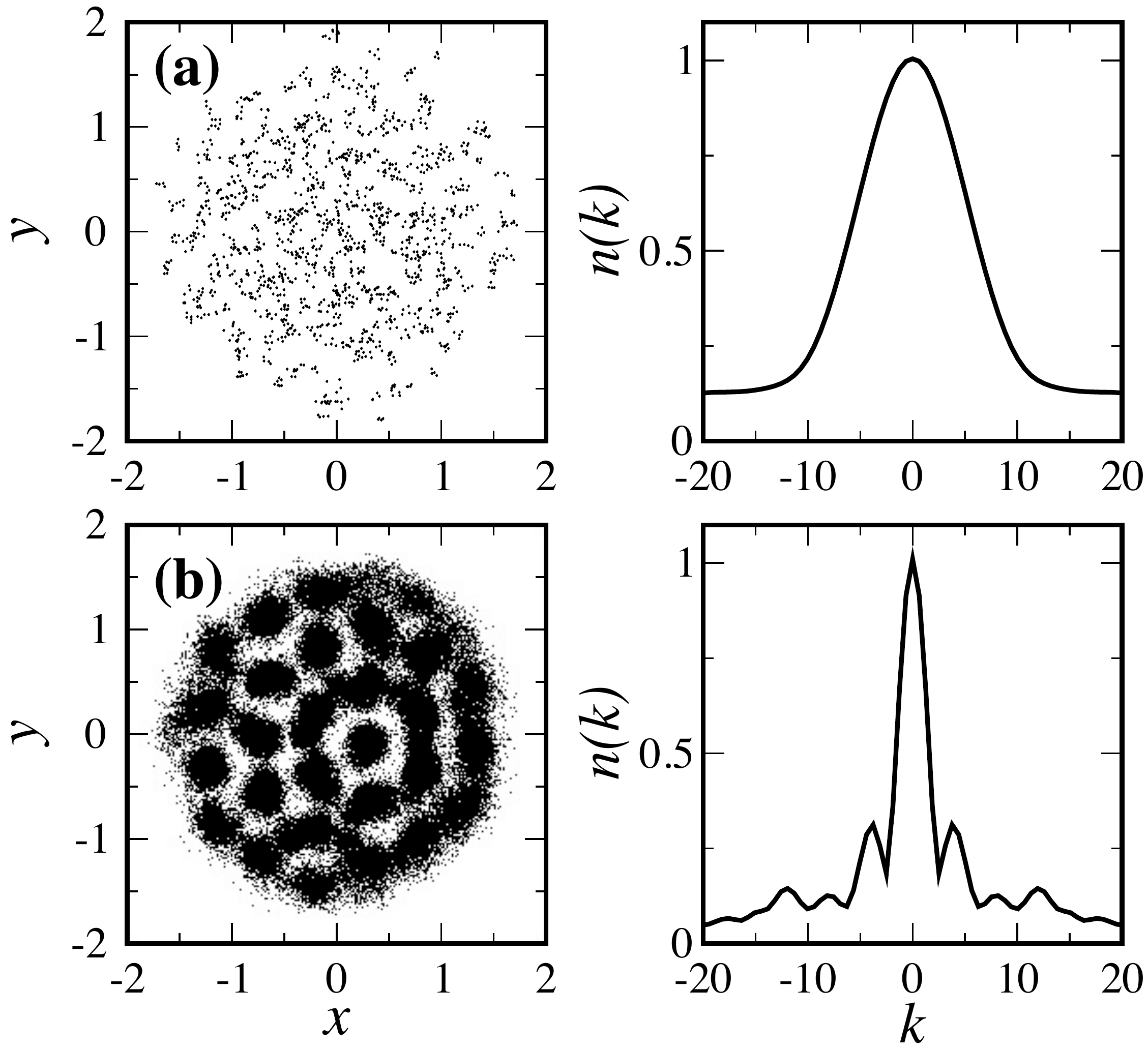}
\caption{(Color online) {\it Left panels}: Monte Carlo snapshots of a system of $N=400$ particles, interacting via the potential \eqref {pot1}, confined in a harmonic trap of strength $\Gamma=500\ \epsilon_\circ$, at the two temperatures $T$=100 $\epsilon_\circ$ (a) and $T$=0.5 $\epsilon_\circ$ (b). {\it Right panels}: corresponding momentum distributions, all normalized to unity for comparison purposes. The value of $R_c$ in this case is 0.3. The development of secondary peaks at low temperature signals the occurrence of a supersolid phase.}
\label{trap}
\end{center}
\end{figure}
The results discussed so far pertain to numerical simulation of the system described by Eq. \eqref{ham2} in its bulk phase. However, in any experiment aimed at probing the physics of such a system, the assembly of particles must necessarily be finite (a few thousand particles is a typical number for current experiments with cold dipolar atoms), confined by an external potential. In order to enable a direct comparison with possible future experiments, we have performed simulations of the same system spatially confined {\it in-plane} by a harmonic trap, i.e., the term $\Gamma \sum_i {\bf r}_i^2$ is added to Eq. (\ref{ham2}), $\Gamma\equiv m\omega^2/2$ being the strength of the trap.
\\ \indent
Fig. \ref{trap} shows typical many-particle configurations of a trapped system comprising $N$=400 particles, at two different temperatures. Also shown are the associated momentum distributions $n(k)$, which are obtained by Fourier transforming of the {\it spherically and translationally averaged} one-body density matrix, computed by Monte Carlo.\\ \indent
Here too, droplets with a well-defined average number of particles form, and organize themselves on a triangular lattice.
Correspondingly, the momentum distribution, which is directly observable experimentally by time-of-flight measurements \cite{greiner}, develops a sharp central peak, with additional structure on its sides.  The secondary peaks correspond to oscillations in the one-body density matrix, in turn reflecting particle tunnelling to adjacent droplets. They are therefore connected to the appearance of the supersolid phase, as explained above.

Summarizing, accurate numerical simulations of a system of dipolar particles interacting via a potential softened at short distance, reveal the existence of a low temperature crystalline phase of superfluid droplets. This phase turns superfluid (supersolid) at $T\to 0$ through a mechanism of tunnelling of particles between adjacent droplets. At higher density this tunnelling is suppressed and an insulating droplet crystal occurs, a phase which has not previously been predicted. The interaction that underlies such intriguing, until now unobserved physical behaviour, can be realized with dipolar atoms in the dipole-blockade regime. 

A comment is on order, concerning the dependence of the results on the particular form of potential utilized here, namely Eq. \eqref {pot1} with its abrupt, sharp cutoff at $r=a$. First off,  the superfluid droplet crystal phase does not crucially depend on the dipolar form of the interaction at long distances. Indeed, it is also observed in our simulations for a van der Waals-like potentials (i.e., $v(r)$ $\sim$ $r^{-6}$).
Secondly,
we have obtained qualitatively similar results with different model potentials, featuring a more realistic ``flat" region at short distances, as well as a smoother merge of long- and short-range behaviours. 
For example, we considered the potential
$V(r)=D/\left(a^3+r^3\right)$, which is naturally realized in a cold gas of alkali atoms by weakly dressing the groundstate $|g\rangle$ of each atom with an excited Rydberg state $|r\rangle$ with a large dipole moment $d$, in the kDebye range~\cite{Pupillo10}. 
\\ \indent
While several dressing schemes are possible~\cite{Henkel10,Honer10}, here we consider $|r\rangle$ as the lowest-energy state of a Rydberg manifold with principal quantum number $n$ for an atom in the presence of a homogeneous electric field $F<F_{\rm IT}$ in the linear Stark regime, with $F_{\rm IT} \propto 1/n^4$ the Inglis-Teller limit. For a laser with (effective) Rabi frequency $\Omega$ and red detuning $|\Delta| \gg \Omega$, the dressed groundstate reads $|\tilde g\rangle \sim |g\rangle -(\Omega/2\Delta)|r\rangle$, and thus $D \propto  (\Omega/\Delta)^4  d^2$, with $d \propto n^2$,  while the cutoff $a\simeq (d^2/ \hbar |\Delta| \epsilon_0)^{1/3}$ arises because of the  Rydberg-blockade mechanism~\cite{Lukin2001}. Spontaneous emission rates $\gamma_r$ from $|r\rangle$ are strongly reduced to values of at most $\gamma \simeq (\Omega/\Delta)^2 \gamma_r$. Observability of the phases above benefits from large values of $\epsilon_\circ$, and in particular of the ratio $\epsilon_\circ/(h \gamma) = [\Omega^2/(4 \pi |\Delta| \gamma_r)]R_{\rm C}^3$, favoring comparatively small values of $r_\circ$~\footnote{2D scattering requires $r_\circ >  a_\perp$, with $a_\perp$ the transverse harmonic oscillator length, in the tens of nm range.}. For example, for $^{87}$Rb atoms in an electric field $F=25$kV/m, with $n=20$, $\Omega/2 \pi = 50$MHz and $|\Delta|/2\pi = 3$GHz, we obtain $d$ = 1450 Debye, $a\simeq$ 400 nm, $r_\circ \simeq$ 210 nm,  $\epsilon_{\circ}/(h \gamma)\simeq 90$, and $\epsilon_{\circ}/k_{\rm B}\simeq$ 120 nK.
Collective many-body effects in the Rydberg-blockade regime~\cite{Honer10} not described by Eq.~\eqref{ham2} should be negligible provided $ (R_c/r_s)^2 \ll (2\Delta/\Omega)^2$, which is readily satisfied for parameters as in Fig.~\ref{default}.


This work was supported in part by the Natural Science and Engineering Research Council of Canada under research grant 121210893, and by the Alberta Informatics Circle of Research Excellence (iCore), IQOQI, the FWF, MURI, U.Md. PFC/JQI, EOARD (grant FA8655-10-1-3081), NAME-QUAM. Useful discussions with I. Lesanovsky, N. Prokof'ev and T. Pohl  are gratefully acknowledged.

\bibliography{bose}

\end{document}